\documentclass[11pt]{article}
\usepackage{subeqnarray}

\usepackage{graphicx}
\usepackage{epstopdf}
\usepackage{color}
\usepackage{fullpage}
\usepackage{setspace}
\usepackage{charter}

\usepackage{hyperref} 
\hypersetup{backref, colorlinks=true, citecolor=blue, linkcolor=blue}
\newcommand{\rref}[1]{\hyperref[#1]{\ref*{#1}}}

\usepackage[authoryear,round]{natbib}

\begin{document}

\title{An Analysis of North Pacific Subsurface Temperatures Using State-Space Techniques
\author{ 
Cindy Bessey\thanks{Email: {\tt cbessey@fiu.edu}}\\
College of Arts and Sciences\\
Florida International University\\
Miami, FL, 33139 USA\\
\vspace{7pt}
\and
Roy Mendelssohn\thanks{Email: {\tt roy.mendelssohn@noaa.gov}}\\
Environmental Research Division\\
NOAA/NMFS/SWFSC\\
Pacific Grove, CA, 93950 USA}
\date{}
}

\maketitle

\begin{abstract}
North Pacific subsurface temperature data from the Simple Ocean Data Assimilation model at 10m, 50m, 75m, 100m and 150m depths, are analyzed using a combination of state-space decomposition and subspace identification techniques to examine the spatial structure of thermal variability within the upper water column. We identify four common trends from our analysis that display the major broad-scale patterns in the North Pacific over a 47 year period (1958-2004): (1) a basin-wide near-surface warming trend that identifies the mid 1980Õs as a change point from a cooling to a warming trend; (2) a contrasting cooling in the central basin and warming along the coast of North America that began in the early 1970Õs; (3) a cooling along the transition zone and the west coast of North America that becomes dominant around 1998;  (4) and contrasting differences in the subarctic and subtropical gyres displaying differences in processes at each depth. We also provide a detailed analysis of the temperature variability at four chosen locations: 52.5N 142.5W (Gulf of Alaska), 37.5N 172.5W (central basin), 37.5N 137.5W (off coast of California), and 27.5N 137.5W (off coast of Baja California) for both 10m and 150m depths.  These results identify subsurface structure, regional heterogeneity, and they also display important differences and similarities in the patterns of subsurface temperature variability when compared to previously published temperature patterns.
\end{abstract}

\section{Introduction}
\label{intro}
 	The purpose of studying basin-wide sea surface and subsurface temperature data is to provide a detailed description, identify trends, and develop a better understanding of inter-annual and decadal variability throughout the water column, with the ultimate goal of increasing our understanding of ocean dynamics and processes. The understanding of this variability is of importance due to its major impact on regional climate and ecosystems.  Most such studies have focused on sea surface temperature (SST) patterns, and have suggested structural linkages in the ocean between the extra-tropics and the tropics \citep{Zhang1999}, as well as the interactions between the ocean and atmosphere \citep{Philander1992, Levitus2005}.  More specifically, the presence of a decadal scale pattern in temperature variability in the upper water column of the Pacific Ocean has also become well documented in the literature \citep{Graham:1994jo, Miller:1994ph, Trenberth:1994ss, Deser:1996nk, Zhang:1997gn, Huang2001}. These studies have suggested that during approximately 1977 to 1988 sea surface temperatures were relatively cooler in the central North Pacific, and relatively warmer along the western coast of the United States and Canada, with proposed decadal shifts (or Ôregime shiftsÕ) occurring in 1976 and 1989.  This decadal behavior in North Pacific sea surface temperatures has been linked with various ocean and atmospheric conditions and is now termed the Pacific Decadal Oscillation (PDO) \citep{Mantua:1997dp}.  The PDO is used as an index to investigate correlations with numerous species throughout various trophic levels \citep{Beamish1993, Benson2002, Hare:2000et, Polovina19951701}.  
	 
	The years identified as change points, or Ôregime shiftsÕ, will influence our interpretation and understanding of the interaction between physical ocean conditions and the surrounding biota.  And although the 1976 regime shift has become widely accepted, some studies question the ability to detect and predict Ôregime shiftsÕ using such statistical techniques as empirical orthogonal function analysis (EOF) (or principal components analysis (PCA))\citep{ISI:000171261600018, ISI:000188458400001, Newman2007}.  These authors do not seem to be questioning whether rapid change has occurred in the North Pacific water temperatures resulting in regime shifts, but rather whether the methods used can differentiate the variability in the PDO from that of an autoregressive series or Gaussian red noise with stationary statistics. Taking this consideration into account, a more recent suite of papers has used an alternative technique to examine the inter-annual and decadal variability in sea surface and subsurface temperatures \citep{Schwing:1997sh, Mendelssohn:2002jl, Mendelssohn:2003um, Palacios:2004km, Bograd2005}.  These studies have used state-space model techniques which have the ability to decompose the observed data into a nonparametric trend term, a nonstationary seasonal component, an autocorrelated cycle component, and a stationary uncorrelated component.  Using this technique, \cite{Mendelssohn:2002jl} found that subsurface temperatures in the California Current system (CCS) displayed an accelerated warming in 1976, but that this warming actually began around 1972.  Furthermore, this technique also revealed that different geographical regions and depth strata within the CCS have clearly distinct temporal patterns of inter-annual and decadal variability in ocean temperature.  \cite{Bograd2005} conducted a similar analysis of sea surface temperatures in the Gulf of Alaska which likewise revealed a warming trend with a change point in the early 1970s, but also reemphasized the significant spatial heterogeneity across the region. Even without the use of these advanced statistical techniques, some very early studies suggested that the pattern of SST cooling in the central North Pacific occurred around 1969 \citep{Douglas1982}.  Clarification of the contrasting views that a warming trend along the western coast of North America and a cooling trend in the central North Pacific started in the early 1970s, versus 1976, would assist in our ability to understand ecosystem responses to environmental change.  Furthermore, not only is it important to understand when these change points occur, but also, it is important to understand regional differences if we are to fully understand how species respond to environmental changes.  
	 
	In this paper, we use the alternative method of state-space techniques to examine common trends in subsurface temperature data at various depths throughout the entire North Pacific.  Our motivation is to examine the spatial and temporal structure of thermal variability within the upper water column.  Our goal is to identify when important change points occur in the time-series, and to obtain a better understanding of the distinct temporal patterns in inter-annual and decadal variability throughout regions in the North Pacific. We describe the first four common trends of our analysis at various subsurface depths (10m, 50m, 75m, 100m and 150m) for the entire North Pacific, and provide a detailed description for four regional locations.  We relate these trends to previously documented SST trends and compare the change points determined from our analysis to those previously published.

\section{Data and Methods}
\label{sec:DM}
\subsection{Data Source}
\label{sec:DM.DS}
	We used monthly mean subsurface temperature data for the years 1958 to 2004 obtained from Simple Ocean Data Assimilation (SODA) model output (\url{http://atmos.umd.edu/~ocean/}).  These data are on a 0.5$^{\circ}$ latitude by 0.5$^{\circ}$ longitude grid, and were obtained for five standard depths (10m, 50m, 75m, 100m, and 150m).  The datasets for each depth were averaged into 5$^{\circ}$ boxes for our area of interest; 20$^{\circ}$N - 65$^{\circ}$N and 110$^{\circ}$E - 250$^{\circ}$E, which corresponds to the region and resolution used to compute the Pacific Decadal Oscillation (PDO) from sea surface temperature data \citep{Mantua:1997dp}.  We refer to boxes based on the latitude and longitude of the southwest corner, a convention used in the Comprehensive Ocean Atmosphere Data Set (COADS) (\url{http://icoads.noaa.gov}). 
	 
	Although the SODA output is modeled data, it includes assimilation of available subsurface data, and it is one of the few full water column data sets that allow for investigating subsurface thermal variability on basin and decadal scales \citep{Carton2000, Shenoi2005}.  The SODA model (version 1.4.2) assimilates in situ and satellite data from a wide variety of sources including data updates through December 2004 from the World Ocean Database, ship intake measurements, moored hydrographic observations, remotely senses sea surface temperature, real-time temperature profile observations from National Oceanographic Data Center/NOAA temperature archives and observations from Tropical Atmosphere Ocean/Triangle Trans-Ocean Buoy Network (TAO/TRITON) mooring thermistor array and Argo drifters \citep{Carton2008}.  A discussion on the accuracy of the SODA analysis can be found in Carton et al. (2000b) where the authors conducted a comparison study using independent observations.  These authors found that when the SODA model was compared to island tide-gauge time series that the model could explain 30\% of the variance in the frequency band between 5-25 years.   However, the authors state that the SODA model shows clear inconsistencies in regards to predicting thermocline water masses in sufficient volume, realistic mixed layer salinity, simulating the effects of important narrow topographic features, and some errors in the Indian Ocean and Southern Hemisphere.   Nonetheless, numerous published studies have since used SODA output to investigate the variability of physical ocean conditions in various ocean basins \citep{Moon:2004kg, Ashok2004, Shi2007}. 

\subsection{State-Space Analysis}
\label{sec:DM.SSM}

	State-space techniques have been used in several fields of study to decompose time series into a variety of components \citep[see][for a comprehensive treatment]{Durbin:2001wl}.  In the past decade, these techniques have also been used and explicitly documented in oceanographic studies \citep{Schwing:1997sh, Schwing1997b, Mendelssohn:2003um, Bograd2005}. We provide the following review of these techniques, a good introduction summary can be found in  \cite{Commandeur:Koopman:Ooms:2010:JSSOBK:v41i01} and other papers in that special volume.	
	
	For each spatial location and depth in our analysis, we assume that each observation $y(t)$ is the sum of four components

\begin{equation}
      y(t) = T(t) + S(t) + I(t) + e(t), \qquad t=1,\tau, 
\end{equation} 

where, at time $t$, $T(t)$ is the unobserved time-dependent mean-level
(nonparametric trend), $S(t)$ is the seasonal component (zero-mean,
nonstationary and nondeterministic),  $I(t)$ is the irregular term (containing
any stationary autocorrelated or cyclic part of the data), and $e(t)$ is the
stationary uncorrelated component, which can be viewed here as "observation"
error. Piecewise continuous smoothing splines are used to estimate the unobserved components.

The trend term can be viewed as a
unknown function of time, and parameterized as 
\begin{equation}
   \nabla^k T(t) \sim N(0,\sigma_T^2). 
   \label{trend_smooth} 
\end{equation} 
where $ \nabla$ is the lag difference operator, $k$ is the degree of the lag which is equal to 1 for all our analysis, $\sim N(0,\sigma_T^2)$ denotes a random variable that is normally distributed with mean $0$ and variance $\sigma_T^2$ which is estimated and controls the smoothness of the estimated trend.  The trend gives a nonparametric estimate of the change of the level (mean) of the series with time.

We constrain the running sum of the seasonal component ($S$) as follows: 
\begin{equation} 
   \sum_{i=0}^{s-1} S(t-i) \sim N(0,\sigma_S^2), \qquad t=1,T, 
   \label{season1_smooth} 
\end{equation} 
where $\sigma_S^2$ is estimated and controls the smoothness of the estimated seasonal component, and $s=12$ for monthly data

The state-space specification of the irregular cyclic term is:
\begin{equation}
\left( 
\begin{array}{c}
   \psi_{t}  \\
   \psi_{t}^{*} 
   \end{array}\right) =
   \rho \left( 
\begin{array}{rr}
   \cos \lambda_{c} &  \sin \lambda_{c} \\
    -\sin \lambda_{c} &  \cos \lambda_{c} 
   \end{array}\right)
   \left( 
\begin{array}{c}
   \psi_{t-1}  \\
   \psi_{t-1}^{*} 
   \end{array}\right) +
   \left( 
\begin{array}{c}
   \kappa_{t}  \\
   \kappa_{t}^{*} 
   \end{array}\right), \qquad t=1,\ldots,T,
\end{equation}
where $\psi_{t}$ and $\psi_{t}^{*}$ are the states, $\lambda_{c}$ is the
frequency, in radians, in the range $0 <
\lambda_{c} \leq \pi $, $\kappa_{t}$ and $\kappa_{t}^{*}$ are two mutually
uncorrelated white noise disturbances with zero means and common variance
$\sigma_{\kappa}^{2}$, and $\rho $ is a damping factor. The damping factor
$\rho$ in (1) accounts for the time over which a higher amplitude event
(consider this to be a "shock" to the series) in the stochastic cycle  will
contribute to subsequent cycles. A stochastic cycle has changing amplitude and
phase, and becomes a first order autoregression if  $\lambda_{c}$ is $0$ or
$\pi$. 
Moreover, it can be shown that as $\rho \rightarrow 1$, then
$\sigma_{\kappa}^{2}
\rightarrow 0$ and the stochastic cycle reduces to the stationary deterministic
cycle:
\begin{equation}
\psi_{t}= \psi_{0}\cos\lambda_{c}t + \psi_{0}^{*}\sin\lambda_{c}t, \qquad
t=1,\ldots,T.
\end{equation}

The observation errors are assumed to be zero mean, independent, and identically distributed as:
\begin{equation}
  e(t) \sim N(0,\sigma_e^2), \qquad t=1,T.
   \label{eq:errorTerm} 
\end{equation} 

	Maximum likelihood estimates of the unknown parameters in the state space model are obtained from the prediction error decomposition which is computed using the output of the Kalman filter \citep{Kitagawa:1984sp, Kitagawa:1985hc}.  A Kalman smoother is then used to compute the mean and variance of smoothed variables. The analyses were performed using the Finmetrics package for S-Plus \citep{zivot2006modeling}, of which the state-space portion is based on the SsfPack software developed by \cite{ssfpack}. These procedures were independently applied to each 5$^{\circ}$ box of data at each depth. Additional references of these procedures are provided by \cite{Kita:Gers:smoo:1996}, \cite{Harvey:1989bx}), \cite{Durbin:2001wl}, and \cite{zivot2006modeling}. 
	
\subsection{Estimation of Common Trends using Subspace Identification Techniques}
\label{sec:DM.subspace}

	The focus of this study is on the underlying long-term thermal trends in the north Pacific.  Ideally,  multivariate Òcommon trendsÓ  \citep[][chapter 8]{Harvey:1989bx} and \citep[][section 3.2.2]{Durbin:2001wl} would be estimated by the same procedure used for the univariate models.  However, this is impractical for problems of this size.  Instead, subspace identification techniques \citep{Aoki:1990nu, Aoki:1997kl}, which provide an approximate estimate to the state-space model, were used to estimate common non-parametric trend terms of the time series produced from the univariate state space analysis.  For each time series, the estimated seasonal ($S(t)$) and irregular terms ($I(t)$) were subtracted from the observed series, leaving a partial residual of the trend ($T(t)$) plus error ($e(t)$) terms.  These partial residual series were then analyzed using AokiÕs (1990) state-space modeling approach.  AokiÕs method is a subspace identification method developed in linear systems theory and time series analysis for finding minimum realizations of multivariate systems.  This reduced dimensions model captures the overall dynamics in the data series, not just the lag-$0$ covariance, using a limited number of components.  These methods can be equated to estimating cointegration and common trends as defined in the econometric and time series literature \citep{Aoki:1997kl}.  The subspace identification method approximates the Hankel matrix  that is formed as the covariances between the past and future of the series (using a lag(lead) of one).  A singular value decomposition of the resulting Hankel matrix is used to determine the number of underlying (non-orthogonal) components in the state vector needed to adequately represent the dynamics of the series. Estimates of all the system matrices in the state-space model can be derived from the singular value decomposition \citep{Aoki:1990nu}.  Subspace identification methods produce estimates close to the true maximum likelihood estimates \citep{Smith2000}, and can be viewed as a computationally more efficient algorithm for large-scale models that produce good approximations to true maximum likelihood estimates

	Estimates of the underlying component series (referred to as common trends throughout this paper) and the Òloading matrixÓ (the observation matrix in the state-space model) can be obtained from the reduced rank matrices produced by the singular value decomposition.  These methods are comparable to the eigenvectors (spatial patterns) and eigenmode amplitudes (time series) from an Empirical Orthogonal Function (EOF) analysis.  However, state-space methods have advantages over EOF analyses.  Time dependence is explicit in state-space techniques, whereas an EOF analysis does not account for time dependence; EOF analysis is invariant to permutations in the rows and columns of the data matrix \citep[for formal proof see][]{Roweis:1999bs}. Also, with our applied method, the underlying states estimated by the reduced rank model need not be orthogonal.  Orthogonal restrictions in EOF analyses can ÒsmearÓ certain types of space-time dynamics, such as when there are propagating waves or other changes in phase relations.  Finally, state-space models permit statistical comparison of different models or assumptions of what is occurring in the data, as well as statistical tests for outliers and change points in any of the components in the models.

\subsection{Modeling Approach}
\label{sec:DM.approach}
	
	As indicated in the previous sections, our analysis required a two-step approach.  First, univariate models and common trends were estimated for each depth at each location; a total of 252 locations (the number of 5$^{\circ}$  boxes with data between 20$^{\circ}$N - 65$^{\circ}$N and 110$^{\circ}$E - 250$^{\circ}$E) each at five depths (10m, 50m, 75m, 100m, and 150m).  Second, a common trend analysis was done for the combination of all 252 time series at each separate depth.  Based on a visual inspection, it was found that the first four components from this common trend analysis reproduced the main features of the estimated univariate subsurface temperature trends.  As the trends estimated by the state-space model are the ÒbestÓ estimate of the mean level for a given time period, change points are identified objectively at points where the slope of the trend term changes (change of sign or significant inflection), but with subjective consideration that the change in slope of the trend occurs over a long enough period to be of physical significance.
	
	Note that the spatial maps provided are the dynamic Ôfactor loadingsÕ from the observation matrix in the Kalman filter representation of the common trends.  One exception is the map of the first common trend, which shows correlations instead of factor loadings because the first common trend contains a scale effect in the factor loadings Ð i.e. the series is not zero mean.  The y-axis of our common trend graphics displays temperature ($^{\circ}$C) on a relative scale, and uses the minimum temperature for the 47 year time series as a reference point of zero. However, in order to obtain the temperature for a particular location, one would multiply the actual common trend temperature by the factor loading values for that particular location.  We provide text indicating the averaged temperature changes for the time period analyzed for various areas throughout the North Pacific basin. As well, we provide additional figures containing the actual temperature change (standardized over the 47 year time period to have mean zero) for four specific locations at two independent depths; 52.5N 142.5W (Gulf of Alaska), 37.5N 172.5W (central basin), 37.5N 137.5W (off coast of California), and 27.5N 137.5W (off coast of Baja California) at both 10m and 150m depths (Figure~\ref{fig:common1}; black boxes denoted a,b,c,d).  These additional graphics display the extent to which each common trend accounts for the overall univariate trend, and provide a visual justification for our use of four common trends.  Finally, to further investigate the thermal stratification of the upper water column, we compute the temperature difference between the 10m and 150m depth series at each of our four locations, using the temperatures obtained by combining common trends 1 through 4.  
	
\section{Results}
\label{sec:R}

\begin{figure}
  \noindent  \includegraphics[width=6in]{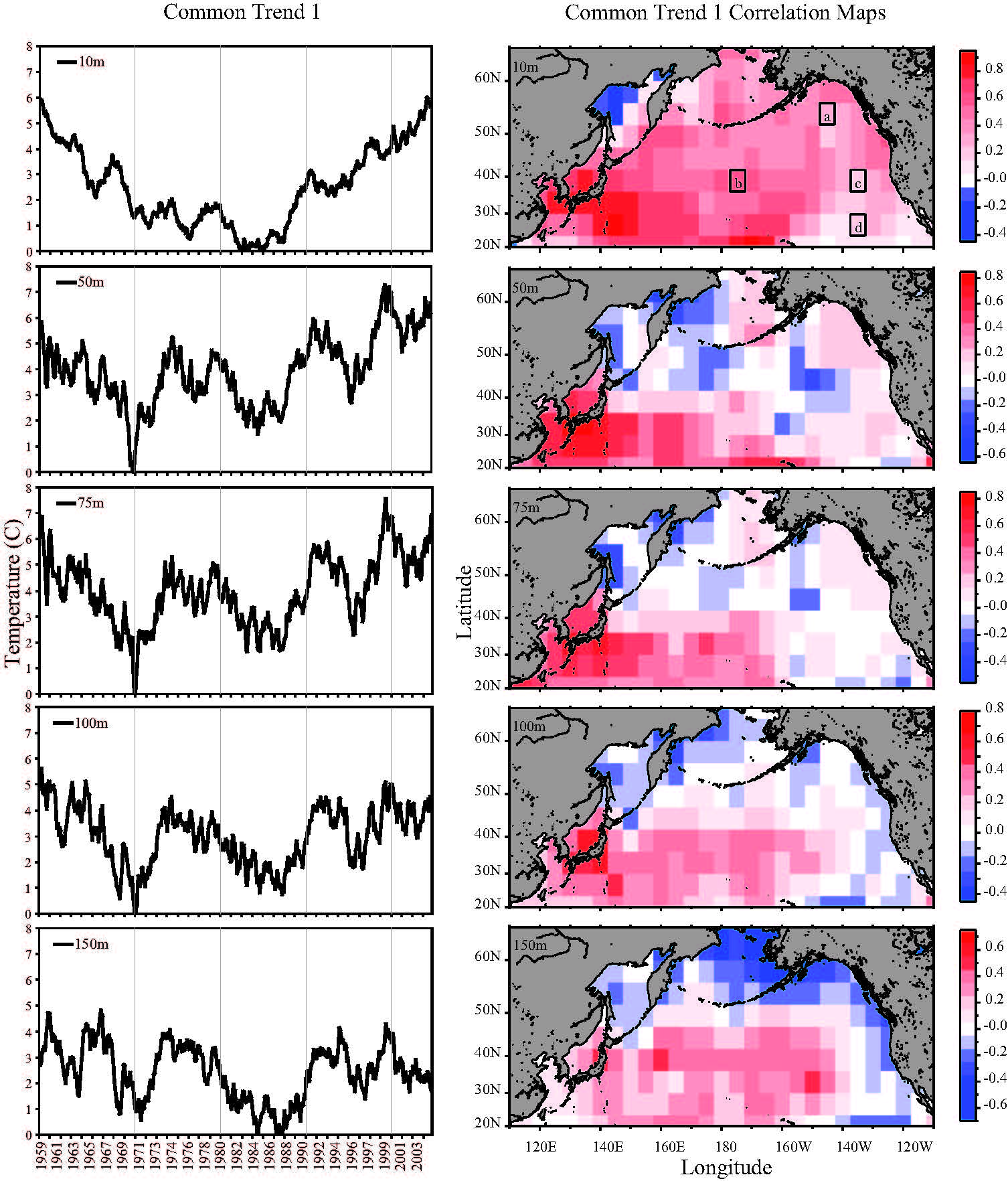}
   \caption{ Common trend 1 of our state-space analysis of SODA subsurface temperature data at 10m, 50m, 75m, 100m, and 150m depths.  The     corresponding correlation patterns between the univariate trend and common trend 1 are shown opposite each common trend 1 graphic.  Letters a, b, c,     and d, represent the four locations chosen for a detailed trend analysis.  The y-axis displays temperature (¡C) on a relative scale, and uses the minimum   temperature for the 47 year time series as a reference point of zero.}
  \label{fig:common1}
\end{figure}

	The first common trend of the subsurface temperature data at 10m depth shows an overall decrease until approximately 1984, followed by an overall increase that continues until the end of the time series (Figure~\ref{fig:common1}).  Similarities to the 10m depth series are also seen in the first common trend in each additional depth series (50m, 75m, 100m, and 150m), which all identify the mid 1980Õs as a change point from a decreasing to an increasing trend.  However, the first common trend of each subsurface temperature series displays a sharp increase during the early 1970Õs that is not seen in the first common trend of the 10m depth series.  Additionally, from the 1990Õs to 2005, the 100m and 150m depth series are relatively level compared to the 10m depth series.  The correlations between the univariate trend and the first common trend for the 10m subsurface data show positive correlations throughout the basin, being particularly strong around the Kuroshio Current region (Figure~\ref{fig:common1}).  This trend represents an increase of approximately 0.6$^{\circ}$C in this area, an increase of approximately 0.3$^{\circ}$C along the coast of North America, and implies a basin-wide near-surface warming trend over the past 45-years.  Likewise, the Kuroshio Extension region also shows positive correlations for the 50m, 75m, and 100m subsurface data, with the maximum correlation shifted slightly to the east at depth, reflecting an increase of 0.5$^{\circ}$C at the 150m.  Correlation for the 10m data contrasts with the deeper data around the Bering Sea and along the west coast of North America, where positive correlations at 10m gradually change to negative correlations by 150m.  This implies significant long-term trends in upper-ocean stratification in these regions.
	
\begin{figure}
  \noindent  \includegraphics[width=6in]{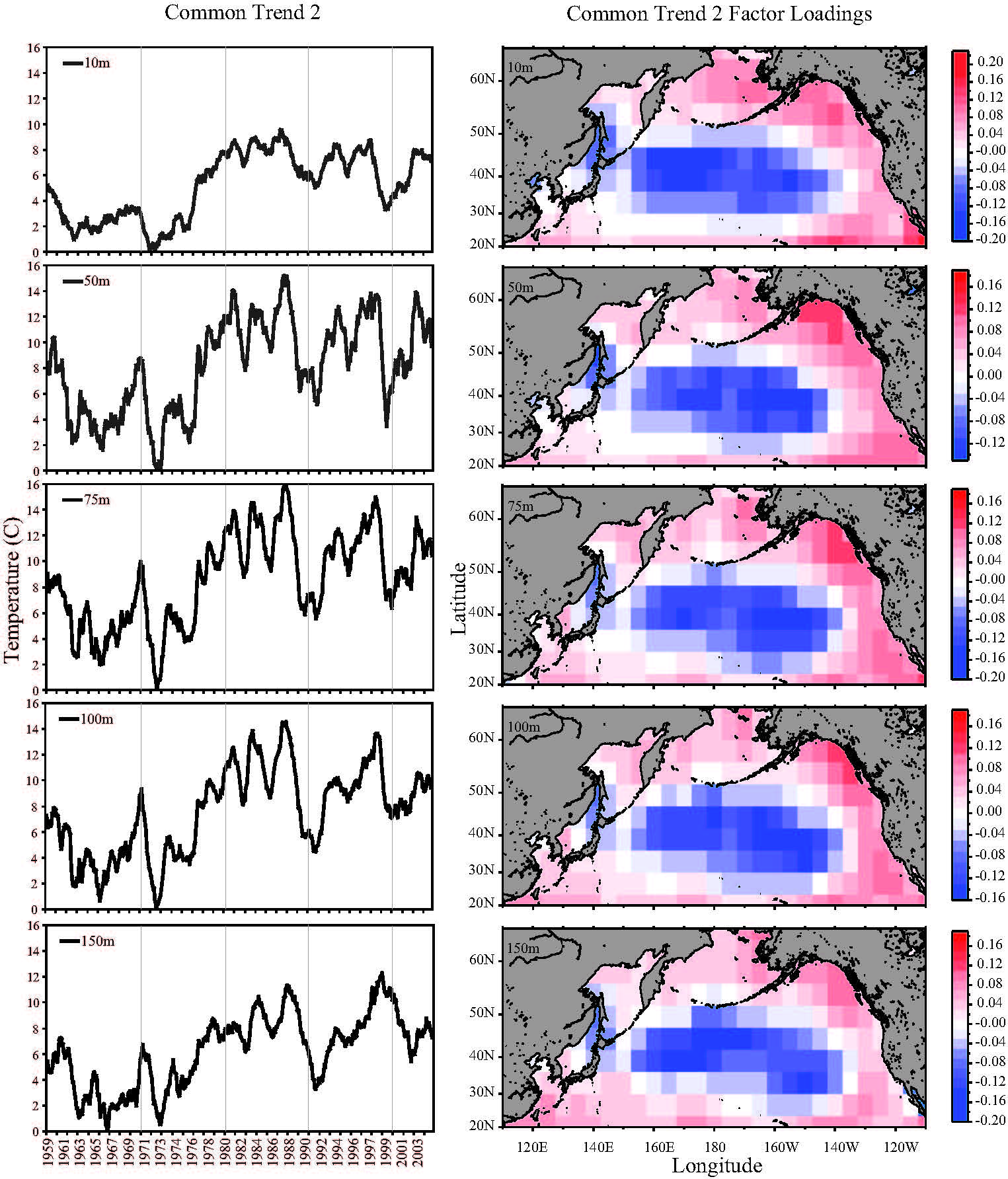}
   \caption{ Common trend 2 of our state-space analysis of SODA subsurface temperature data at 10m, 50m, 75m, 100m, and 150m depths.  The corresponding factor loading patterns are shown opposite each common trend 2 graphic.  The y-axis displays temperature (¡C) on a relative scale, and uses the minimum temperature for the 47 year time series as a reference point of zero.}
  \label{fig:common2}
\end{figure}

	The second common trend of the subsurface temperature data is similar for all depths (Figure~\ref{fig:common2}).  Common trend two displays a sharp decrease during the early 1970s for all depth series.  This is followed by an increasing trend which again is seen in all depth series by 1973, and accentuated by an accelerated increase around 1976.  Several sharp decreases/increases are seen in 1982/83, 1987/88, the early 1990Õs, 1998 and 2002, which are identifiable as El Ni\~{n}o years.  At each depth, the factor loadings are positive in the Bering Sea and extending down the west coast of North America, and negative around the central North Pacific gyre, indicating warming along the west coast concomitant with cooling of the central basin (Figure ~\ref{fig:common2}).  At both the 10m and 150m depths, this represents an overall decrease of approximately 1.0$^{\circ}$C in the central basin, and an average increase of approximately 0.7$^{\circ}$C for the entire west coast of North America at the 10m and 0.07$^{\circ}$C at the 150m depths.  The variability described by the second common trend displays similar spatial-temporal patterns at all depths. However, an exception occurs during 1999-2002, when the trend for the 10m and 150m depth series diverge, indicating reduced stratification.
	
\begin{figure}
  \noindent  \includegraphics[width=6in]{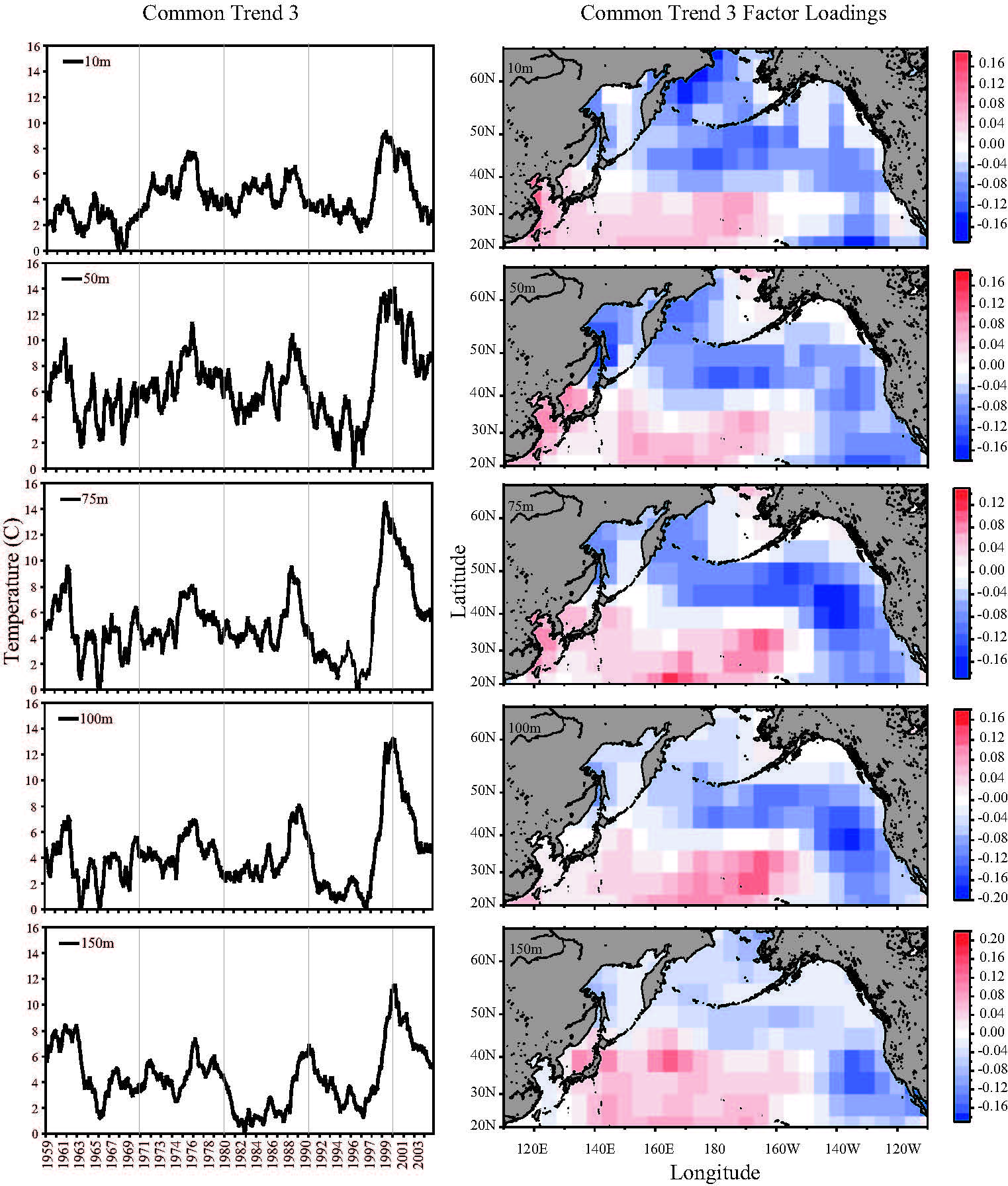}
   \caption{ Common trend 3 (see Figure 2 caption for full details).}
  \label{fig:common3}
\end{figure}

	The third common trend of the subsurface temperature data is again similar for all depths (Figure~\ref{fig:common3}).  This trend, although relatively level throughout the entire time series for all depths, is characterized by a sharp peak centered in the late 1990s.  Factor loadings of the third common trend for all depths are negative along the transition zone and down the west coast of North America (Figure~\ref{fig:common3}), and positive in the southwestern North Pacific.  This represents an average decrease of approximately 0.4$^{\circ}$C for the 10m depth series along the transition zone and the west coast of North America, but an average decrease of approximately 0.7$^{\circ}$C at each subsurface depth in these areas.  
	
\begin{figure}
  \noindent  \includegraphics[width=6in]{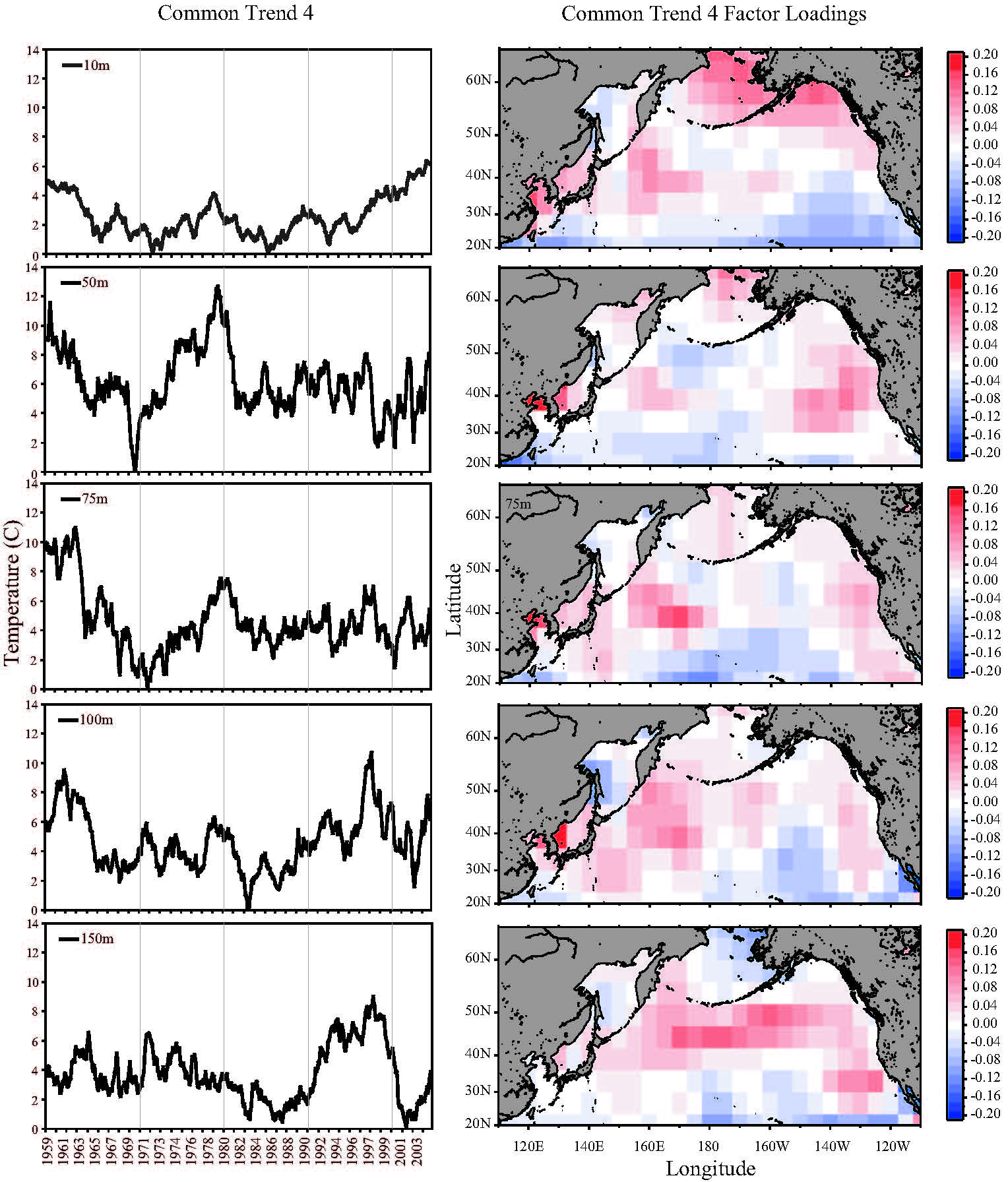}
   \caption{ Common trend 4 (see Figure 2 caption for full details).}
  \label{fig:common4}
\end{figure}

	The fourth common trend of the subsurface temperature data at 10m shows an initial decrease until 1972, followed by an increase until 1979, then a decrease until 1986, and ending with an overall increase until the end of the time series (Figure~\ref{fig:common4}).  This contrasts the pattern seen at the 50m and 75m depths, which shows a similar decrease until 1970, followed by a sharp increase/decrease from 1979-1981, but then remains relatively level until the end of the time series.  Furthermore, the fourth common trend for the 100m and 150m depths display a gradual decrease until 1984, followed by an increase until 1998, ending with a decreasing trend that displays a pronounced decrease spanning 1999 to 2001. The factor loadings for the 10m data series are positive north of California along the coast of North America and throughout the Gulf of Alaska and Bering Sea, and are negative in the southeastern North Pacific (Figure~\ref{fig:common4}).  The average increase in temperature in the Bering Sea is approximately 0.5$^{\circ}$C at the 10m depth.  At 150m depth, the factor loadings are positive along the majority of the coast of North America (Figure~\ref{fig:common4}), with positive factor loadings extending west through the transition zone.  The average approximate temperature increase of the transition zone is 0.8$^{\circ}$C at the 150m depth.  
	
\begin{figure}
  \noindent  \includegraphics[width=6in]{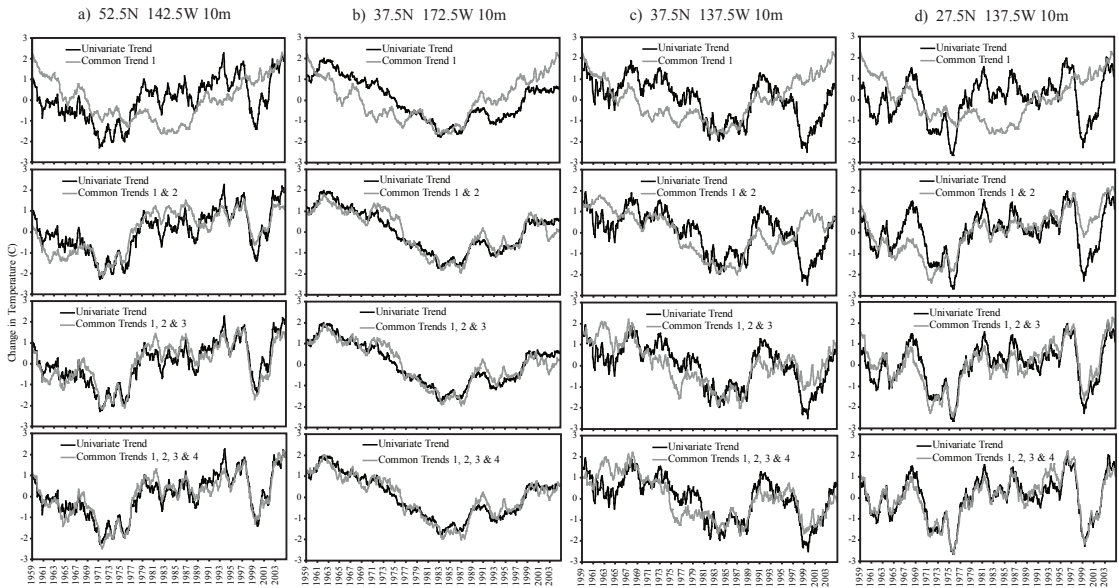}
   \caption{ A comparison between the univariate trend and common trend 1, and the addition of each subsequent common trend at the 10m depth for four chosen locations: a) 52.5N 142.5W; Gulf of Alaska, b) 37.5N 172.5W; central basin, c) 37.5N 137.5W; off the coast of California, and d) 27.5N 137.5W; off the coast of Baja California. The y-axis is the actual change in temperature for a particular time point (standardized over the 47 year time period to have mean zero).}
  \label{fig:comparison10m}
\end{figure}

\begin{figure}
  \noindent  \includegraphics[width=6in]{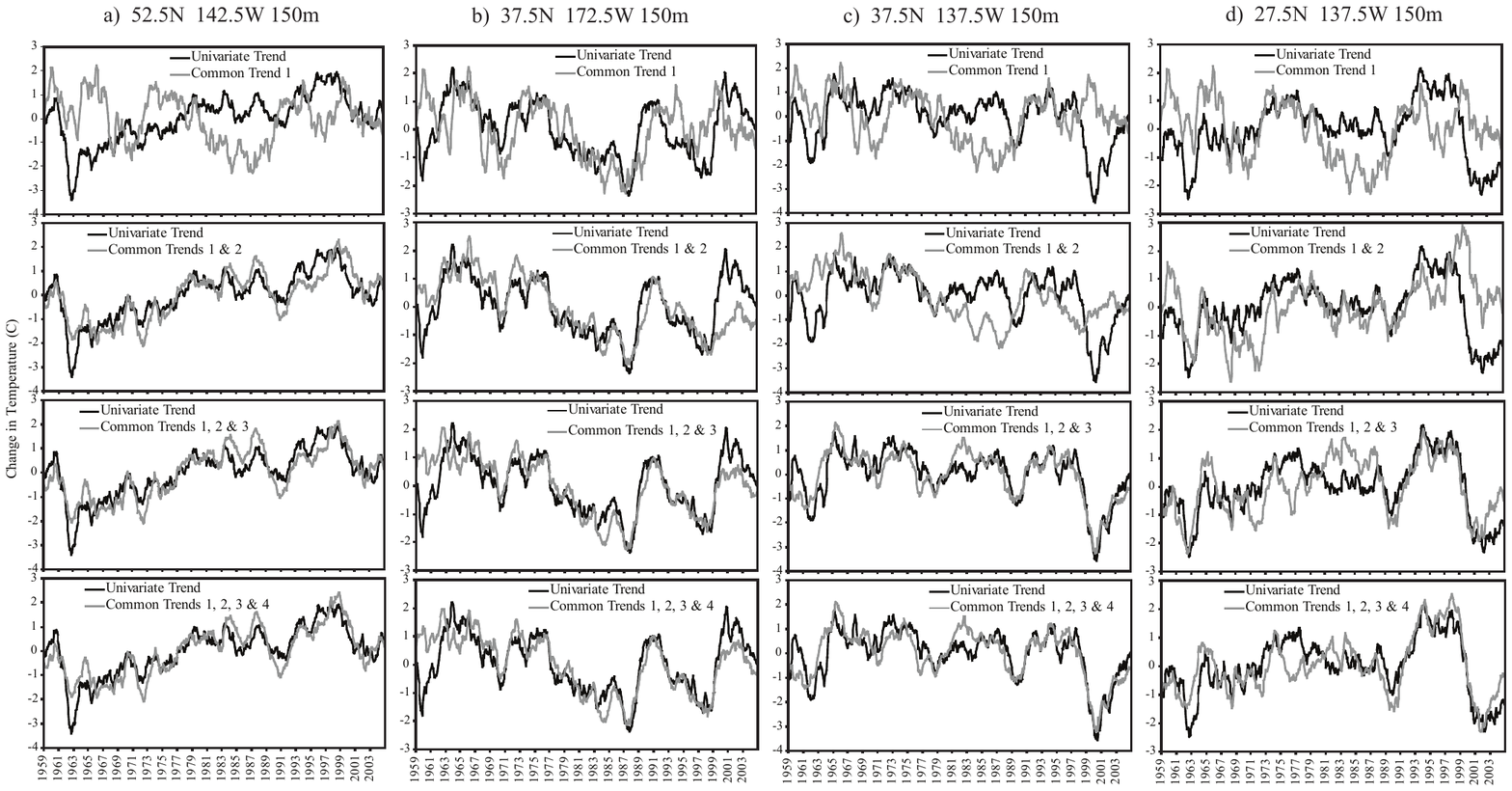}
   \caption{ A comparison between the univariate trend and common trend 1, and the addition of each subsequent common trend at the 150m depth for four chosen locations (see Figure 5 caption for location details).}
  \label{fig:comparison150m}
\end{figure}

\begin{figure}
  \noindent  \includegraphics[width=6in]{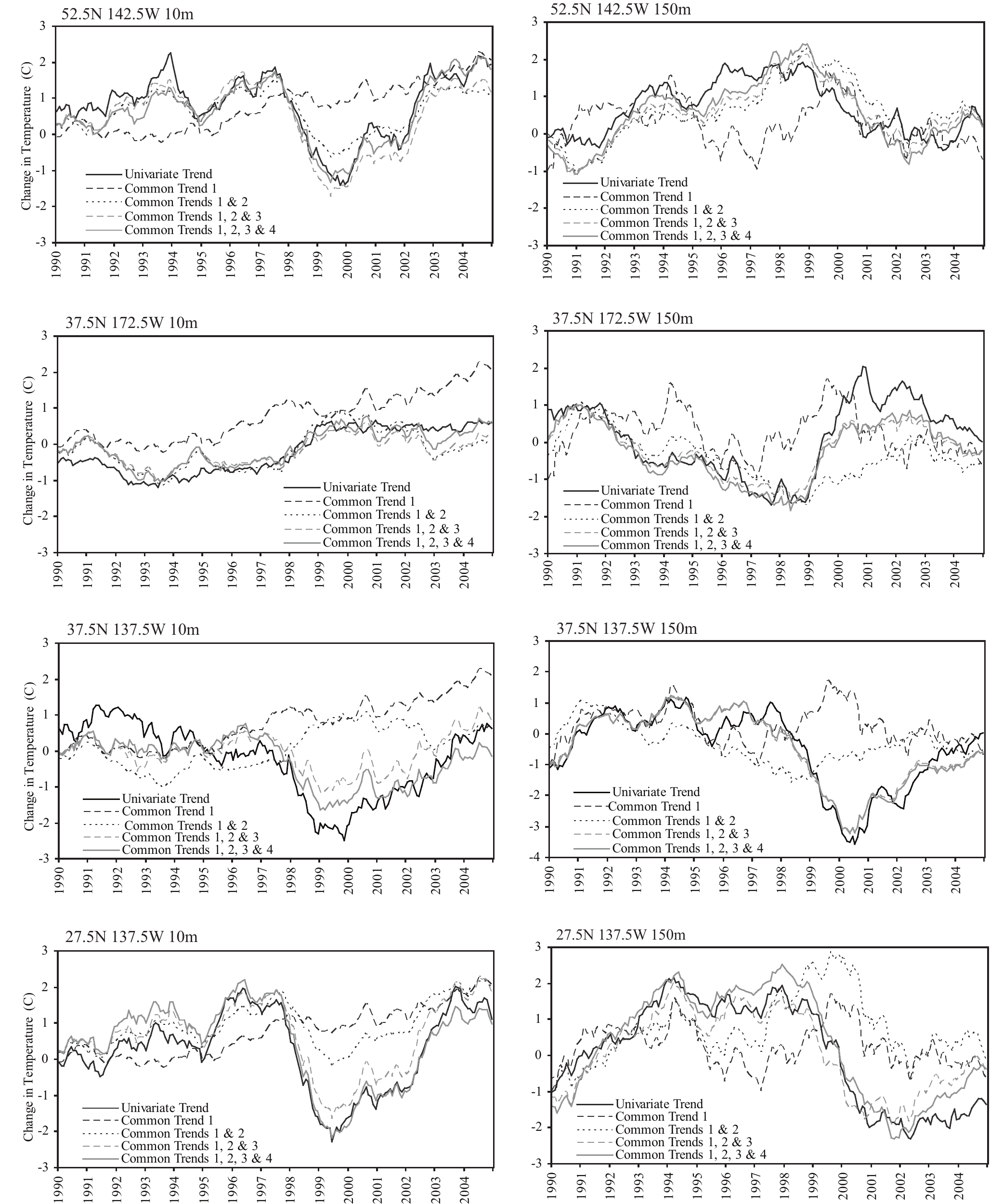}
   \caption{ A comparison between the univariate trend and common trend 1, and the addition of each subsequent common trend at both the 10m and 150m depth for four chosen locations from 1990-2005 (see Figure 5 caption for location details)}
  \label{fig:comparison10m150m}
\end{figure}

  	We chose four specific locations to further display the actual temperature changes and the extent to which each common trend is able to reconstruct the univariate trend.  The 10m and 150m depth series for each of our specific locations are displayed in Figure~\ref{fig:comparison10m} and Figure~\ref{fig:comparison150m}, respectively.  Specifically, at the 10m depth in the Gulf of Alaska  (52.5N 142.5W), the combination of common trends 1 and 2 is reasonably able to reconstruct the univariate trend (Figure~\ref{fig:comparison10m}a).  However, without the addition of common trends 3 and 4, the behavior of the univariate trend during the early 1960Õs would not be reproduced.  Our central basin location (37.5N 172.5W) at the 10m depth is reproduced with just the addition of common trends 1 and 2 (Figure~\ref{fig:comparison10m}b), but the addition of common trends 3 and 4 are required at the 150m depth to capture the rapid increase that occurs around 1999, followed by a decrease around 2002 (Figure~\ref{fig:comparison150m}b).   Our locations off the coasts of California (37.5N 137.5W) and Baja California (27.5N 137.5W) provide evidence for the importance of common trend three.  The combination of common trends 1 and 2 fail to capture the univariate trend behavior during the late 1990Õs (specifically, the 1998 decrease) and onward at both locations for both the 10m and 150m depths (Figure~\ref{fig:comparison10m}c,d and Figure~\ref{fig:comparison150m}c,d).  However, when the third common trend is used the univariate trends are then more reasonably duplicated.  Furthermore, off the coast of California at the 150m depth, the combination of common trends 1 and 2 fail to capture the behavior around 1988, and in fact show an increase when the univariate trend is clearly decreasing.  This behavior is captured with the addition of the third common trend.  To emphasize the extent to which each common trend contributes to the overall univariate trend, we have enlarged a 15 year period (1990 Ð 2005) of these time series (Figure~\ref{fig:comparison10m150m}).  Again, our locations off the coasts of California and Baja California at the 10m depth both display how common trend four enhances the ability to replicate the univariate trend during the late 1990Õs and early 2000Õs.   However, off the coast of California at 150m depth, the fourth common trend adds very little to the overall ability to duplicate the univariate trend.  These figures enable the reader to determine the approximate temperature change that would not be accounted for depending on the number of common trends used.  For example, at 37.5N 137.5W at the 10m depth during 1999, if only common trends one and two were used, a temperature difference of almost 3$^{\circ}$C would be neglected compared to the univariate trend.  However, with the addition of common trend three, the difference is reduced to about 1$^{\circ}$C, and further reduced to 0.5$^{\circ}$C with the addition of common trend four.
	
\begin{figure}
  \noindent  \includegraphics[width=6in]{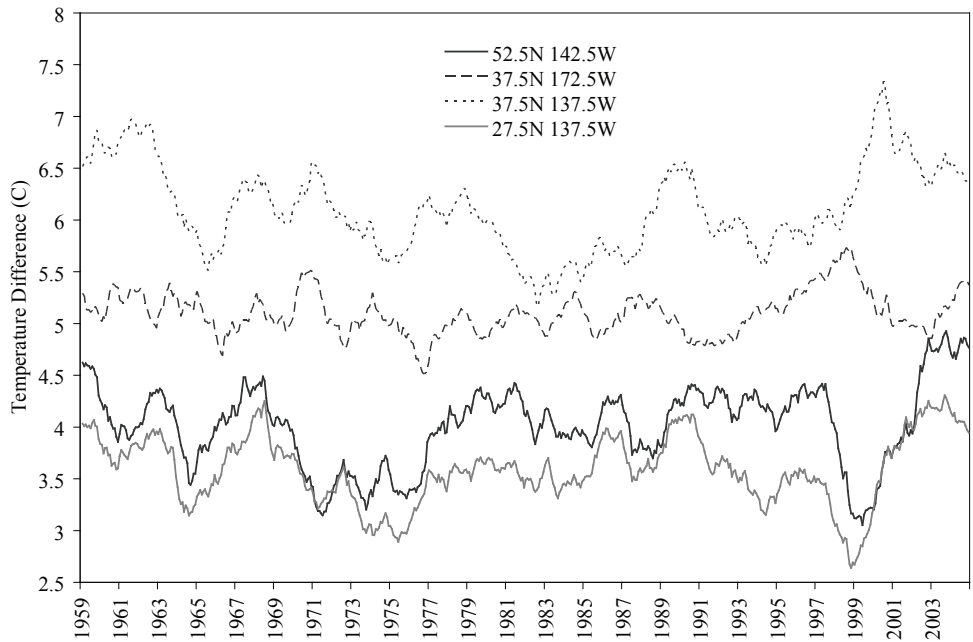}
   \caption{ The difference in temperature between the 10m and 150m depth series using the temperatures obtained by combining common trends 1 through 4 for four chosen locations (see Figure 5 caption for location details).}
  \label{fig:diff}
\end{figure}

	Using the combination of common trends one through four, we subtracted the 150m depth temperatures from the 10m depth temperatures to investigate thermal stratification (Figure~\ref{fig:diff}).  The locations off the coast of California and the central basin display the most stratification, as identified by their greatest difference in temperature from the 10m to the 150m depth.  While sharp increases or reductions in stratification are not noticeable in the central basin, they are evident in each of the remaining three locations.  A sharp reduction in stratification is observed during 1963/1964, 1980/1981, 1991 and 2000, while a sharp increase is observed during 1999 off the coast of California.  The Gulf of Alaska and also the location off the coast of Baja California display similar trends in stratification.  These locations display a sharp reduction in stratification during 1964, 1969/1970, 1987 and 1998, and a sharp increase in stratification during 2000. 
	
\section{Discussion}
\label{sec:Disc}
The main objective of our study was to identify the timing of change points which could reflect broad climate shifts, and to investigate patterns in the inter-annual and decadal variability of subsurface temperatures throughout the North Pacific.  Our analysis describes common trends in the SODA subsurface temperature data at various depths (10m, 50m, 75m, 100m and 150m) using state-space techniques.  This examination of the spatial and temporal structure of thermal variability within the upper water column revealed some similarities in trends, and some differences in change point timing, when compared to previously documented indices and temperature trends.  Here we compare our results to earlier studies on North Pacific sea surface and subsurface temperature trends, and we then explore the potential significance of the identified differences. 
  
	 Our analysis suggests that changes in temperature trends for the North Pacific occurred in the early 1970Õs and mid 1980Õs for the 47 year period examined.  The temperature trend displays an overall temperature increase starting in the mid 1980Õs, reflecting an increase of approximately 0.6$^{\circ}$C in the Kuroshio Current area, and an increase of approximately 0.3$^{\circ}$C along the coast of North America.  These results imply a basin-wide near-surface warming trend and suggest that this long-term warming trend is also observed in the upper mixed layer of the North Pacific Ocean.  Furthermore, and more specifically, our results also indicate warming along the west coast concomitant with cooling of the central basin of the North Pacific which started in the early 1970Õs.  This cooling SST in the central Pacific and warming off the coast of North America has been well documented \citep{Graham:1994jo, Miller:1994ph, Deser:1996nk, Mantua:1997dp, Hare:2000et, Suga1993, Levitus2005}.  However, the difference between these studies and ours is in the identification of when these patterns emerged.  Our results suggest that these changes occurred in the early 1970Õs and mid 1980Õs, which contrasts the widely published view of both a 1976 and 1989 Òregime shiftÓ (Hare and Mantua 2000).    The Pacific Decadal Oscillation Index (PDO) is widely referenced in this regards and represents the leading principle component of North Pacific monthly sea surface temperature variability \citep{Mantua:1997dp}.  \cite{Mendelssohn:2003um} conducted a similar analysis of subsurface temperature data specific to the California Current System which produced results (common trend 1) similar to those observed in our current study, with an emphasis on the shift occurring in the early 1970Õs and accelerated during the 1976 period.  Their analysis used historical data obtained from the World Ocean Database, indicating that not only does our analysis reproduce previously observed behavior, but that there is consistency between both the SODA model and observed data.  The identification of change points during the early 1970Õs and mid-1980Õs, compared to the previously published changes of 1976 and 1989 are in themselves significant results.  The accurate identification of important physical changes is critical in our ability to interpret ecosystem responses and identify possible mechanisms of change.  A difference of four years could significantly affect our interpretation of possible mechanisms. 
   
	Furthermore, our analysis also demonstrates how environmental ocean observations can be separated into spatial and temporal modes of variability.  The detailed examples of thermal variability at four chosen locations display the regional heterogeneity of the North Pacific.   The regional heterogeneity observed in our study provides evidence that a simple model of ÒwarmÓ regime and Òcold regimeÓ is not sufficient for the entire North Pacific.  For example, although the combination of common trend 1 and 2 capture the major behaviors of the univariate trend for the Gulf of Alaska and the central basin area, this combination alone would not be sufficient to capture all major behaviors in our California Coast and Baja California locations.  This is an important aspect when relating ocean conditions to fishery population dynamics that vary by basin.  
   
	The third common trend of the subsurface data shows similar spatial and temporal patterns at each depth, and is characterized by an increase around 1998.  This 1998 change is also observed in the sea surface height data and represents a transition to a state characterized by above-average sea level in the Central Pacific and below average sea level adjacent to the North American coast and in the Gulf of Alaska \citep{Cummins2005}.  Likewise, the spatial patterns of trend three are similar to those of the second principal component of an analysis conducted on SST anomalies in the North Pacific; the ÒVictoria ModeÓ \citep{Bond2003}.  Our analysis expands on these previous studies by showing that the trend and patterns seen extend down to a depth of at least 150m.
   
	The factor loadings for the fourth common trend provide evidence that different processes may be responsible for the observed temperature trends at each depth.  We hypothesize that this trend identifies differences between the subarctic and subtropical gyres based on the spatial patterns of the loadings. The spatial pattern of the loadings at 10m depth is positive in the area of the Alaska Current and Bering Sea, contrasted by negative loadings in the area of the Subtropical Gyre.  However, at depth (150m), the spatial pattern of the loadings is predominately positive along the area of the western subtropical Pacific and extends down into the California Current area.  The spatial pattern of the loadings for the additional depths (50m, 75m, and 100m) displayed intermediate configurations.
   
Our analysis documents trends and patterns in the subsurface temperatures of the entire North Pacific, and relates these trends to previously documented SST trends and other oceanographic indices. Our identified trends display consistencies with past analyses and emphasize important change points in ocean temperature structure.  Although we point out similarities to previous indices, our analysis provides examples of differences and extends our current understanding of thermal variability throughout the upper water column.  The continued progress in our understanding of thermal variability throughout the North Pacific will enable us to better understand ocean processes and how the oceanÕs inhabitants respond to this variability. Effective management of marine resources requires a better understanding of the interactions between the changing ocean conditions and their impacts on the surrounding biota.

\newpage

\bibliographystyle{crs}
\bibliography{sodaPDO}

\newpage

\end{document}